\input{aipcheck}

\documentclass[
    ,final            
  ]
  {aipproc}

\layoutstyle{6x9}
\newcommand{\pom}{I\!\! P}
\begin{document}
\vspace*{-0.5in}
\title{Aspects of Diffraction at the Tevatron\\
{\normalsize \it Review and Phenomenological Interpretation of CDF Results on Diffraction}\\
{\normalsize\bf Presented at CIPANP-2003, New York City, 19-24 May 2003}} 
\author{Konstantin Goulianos}
{address={The Rockefeller University\\
1230 York Avenue, New York, NY 10021\\
{\rm (Experimental results are presented on behalf of the CDF Collaboration)}}
}



\begin{abstract}
Results on soft and hard diffraction obtained by the CDF 
Collaboration at the Fermilab Tevatron $\bar pp$ Collider are reviewed 
with emphasis on aspects of the data that point to the underlying QCD 
mechanism for diffraction. The results are interpreted in terms of 
a phenomenological approach in which diffraction is due to 
an exchange of low-$x$ partons subject to color constraints.   
\end{abstract}

\maketitle

\section{INTRODUCTION}
Diffractive interactions between hadrons are characterized by the 
presence of one or more large rapidity gaps in an event.  Processes which 
({\em do not}) incorporate a hard partonic scattering in addition to the 
rapidity gap signature of diffraction are referred to as ({\em soft}) 
hard diffractive. A rapidity 
gap is a region of pseudorapidity\footnote{We use {\em rapidity} and 
{\em pseudorapidity} interchangeably, since in the kinematic region of interest 
in this paper the pseudorapidity of a particle, defined as 
$\eta=-\ln\tan\frac{\theta}{2}$, where $\theta$ is the polar angle, is 
numerically very close to its rapidity, $y=\frac{1}{2}\frac{E+p_L}{E-p_L}$,
where $p_L$ is the longitudinal momentum of the particle.} 
devoid of particles. Rapidity gaps may be formed in  
non-diffractive (ND) interactions by multiplicity fluctuations. However, 
from Poisson statistics, the probability for a ND gap of 
width $\Delta\eta$ is expected to be 
P$(\Delta\eta)=\exp[-\rho\Delta\eta]$, where 
$\rho$ is the average particle density per unit $\eta$.
Thus, ND gaps are exponentially suppressed with increasing $\Delta\eta$. 
In contrast, diffractive gaps do not exhibit such a suppression. 
This aspect of diffraction could be explained if the exchange across 
the gap were a color singlet quark/gluon object with vacuum quantum 
numbers. For historical reasons, this object is referred to as 
Pomeron~\cite{Regge}. In this paper, we briefly review  
the results on soft and hard diffraction reported by the 
Collider Detector at Fermilab (CDF), present new results from Run~II, 
and ``interrogate'' the data to learn about 
the partonic structure and factorization properties of Pomeron exchange.
The information obtained is compared with expectations form the 
``renormalized gap probability'' phenomenological model (RENORM), 
in which the Pomeron is formed from  
the underlying partonic structure of the interacting hadrons 
subject to the color-matching requirements appropriate for 
``vacuum exchange''~\cite{R,corfu}. 

The paper is organized in two sections: soft diffraction and hard diffraction.
For pedagogical reasons, experimental results and RENORM 
model expectations are presented concurrently and conclusions are 
interspersed within the main body of the presentation.

\section{SOFT DIFFRACTION}
\centerline{The following soft $\bar{p}p$ processes have been studied by CDF:}
\begin{center}
\begin{tabular}{lll}
{\bf ND}&Non-Diffractive&$\bar{p}p\rightarrow X$\\
{\bf SD}&Single Diffraction~\cite{sd}&$\bar{p}p\rightarrow \bar{p}+{\rm gap}+X$\\
{\bf DD}&Double Diffraction~\cite{dd}&$\bar{p}p\rightarrow X+{\rm gap}+Y$\\
{\bf DPE}&Double Pomeron Exchange~\cite{idpe}&$\bar{p}p\rightarrow \bar{p}+{\rm gap}+X+{\rm gap}+p$\\
{\bf SDD}&Single $\oplus$ Double Diffraction~\cite{sdd}&
$\bar{p}p\rightarrow \bar{p}+{\rm gap}+X+{\rm gap}+Y$\\
\end{tabular}
\end{center}
\vglue -0.1in
\begin{figure}[h]
\includegraphics[width=0.75\textwidth]{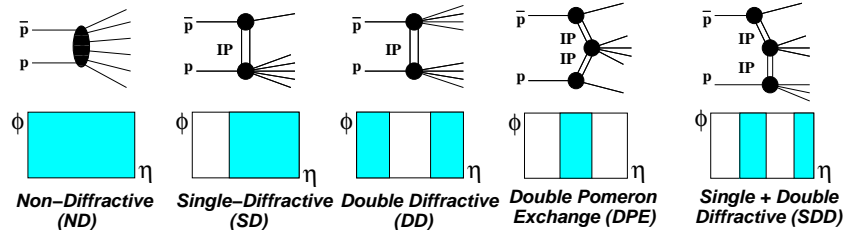} 
\caption{Diagrams and $\eta$-$\phi$ topologies of soft processes studied 
by CDF; the shaded areas are regions where particle production occurs 
and are referred to in this paper as diffractive clusters.}
\label{fig:soft}
\end{figure}

Diffraction has been traditionally treated phenomenologically in the 
framework of Regge theory. The connection of the theory to QCD 
is best seen by expressing cross sections in terms of rapidity 
gap and ``diffractive cluster'' variables, with the latter 
defined as regions of pseudorapidity where particle production 
occurs.
The SDD process, for example, has two rapidity gaps and two diffractive 
clusters,  which we designate, from left to right 
in Fig.~\ref{fig:soft}, as $\Delta\eta_1$, $\Delta\eta_1'$, 
$\Delta\eta_2$ and $\Delta\eta_2'$. The gap $\Delta\eta_1$ can be thought 
of as being formed by the elastic scattering 
between the $\bar{p}$ and the cluster 
$\Delta\eta_1'$, and the gap $\Delta\eta_2$ by the elastic scattering 
between the two diffractive clusters. Each gap is
associated with a four-momentum transfer squared, $t$. There are 5 independent 
variables in SDD: the two rapidity gaps with their associated $t$-values
and the center of the ``floating'' gap (non-adjacent to the $\bar{p}$), 
$\eta_c$.
The Regge theory SDD differential cross section is given by
\begin{eqnarray}
{d^{5}\sigma\over dt_1\,dt_2\,d\Delta\eta_1\,d\Delta\eta_2\,d\eta_c}=
&P_{gap}(t_1,t_2,\Delta\eta_1,\Delta\eta_2,\eta_c)\times
\kappa^2 \times \sigma_{tot}(s')\\
P_{gap}(t_1,t_2,\Delta\eta_1,\Delta\eta_2,\eta_c)=
&C\times F_{\bar{p}}^2(t_1)\times 
\left[e^{(\epsilon+\alpha't_1)\Delta\eta_1}\right]^2\times
\left[e^{(\epsilon+\alpha't_2)\Delta\eta_2}\right]^2\\
\sigma_{tot}(s')=
&\beta(0)^2\;(s')^\epsilon=\beta(0)^2\;e^{\epsilon\ln s'}=
\beta(0)^2\;e^{\epsilon(\Delta\eta_1'+\Delta\eta_2')}
\end{eqnarray}  
where $\beta(0)$ is the $\pom$p coupling at $t=0$, $\epsilon$ and $\alpha'$ 
the parameters of the Pomeron trajectory, $\alpha(t)=1+\epsilon+\alpha't$,  
$\kappa=g^{\pom\pom\pom}/\beta^{\pom p}$ the ratio of the triple-Pomeron to the Pomeron-proton couplings, $s'$ the diffractive cluster sub-energy
defined by $\ln s'=\Delta\eta'=\Delta\eta_1'+\Delta\eta_2'$, and $C$ a constant~\cite{corfu}.
The parameter $\kappa$ has been measured to be 
$\kappa=0.17\pm 0.02$~\cite{GM}.

\paragraph{The QCD connection} There are three factors  in Eq.~(1): 
$P_{gap}$, $\kappa^2$ and $\sigma_{tot}$.
Recalling that the total $\bar{p}p$ cross section is 
$\beta(0)^2\;e^{\epsilon\ln s}$, 
the last factor is identified as the $\bar{p}p$ cross section at the 
diffractive sub-energy squared, $s'$. From the optical 
theorem, the term $e^{\epsilon\Delta\eta'}$ is proportional to the 
forward elastic scattering amplitude at $s'$. The fact that the 
two diffractive clusters are not contiguous does not present 
a conceptual problem in the parton model, in which the amplitude 
is $~\sim e^{\epsilon\Delta\eta_i'}$ for each cluster~\cite{levin} and thus 
the regions $\Delta\eta_1'$ and $\Delta\eta_2'$ add in the exponent. 
The full $t$-dependent parton model amplitude is: 
\begin{equation}
f_{\bar{p}p}(t,\Delta\eta)\propto e^{(\epsilon+\alpha't)\Delta\eta}\;\;\;\;\;\;
{\rm Parton\;Model\;Amplitude}
\end{equation}
Thus, in Eq.~(2), the terms in the square brackets are identified as the
amplitudes for elastic scattering between the diffractive clusters 
on either side of each gap, while $F_{\bar{p}}(t_1)$ is the $\bar{p}$ 
form factor. Finally, the parameter $\kappa$ is identified as 
the color factor required to produce a color singlet exchange; two
such factors are needed in SDD, one for each gap. 

Similar equations can be written for SD, DD and DPE~\cite{corfu}. In all cases 
the cross section factorizes into $P_{gap}(\Delta\eta)$ and 
$\sigma_{tot}(\Delta\eta')$ terms. The predicted shapes of the differential 
cross sections for all four processes agree with the CDF 
data~\cite{dd,idpe,sdd,GM} . However, as seen in Fig.~\ref{fig:softdata}~(a,b),
the $s$-dependence of the SD and DD cross sections is approximately flat
at high energies, contrary to the Regge theory expectation of 
$\sim s^{2\epsilon}$. The culprit for this problem was identified~\cite{R} 
as the normalization of the $P_{gap}(\Delta\eta)$ term, which is obtained 
from the elastic and total cross sections using factorization 
{\em independently} from the normalization of the 
$\sigma_{tot}(\Delta\eta')$ term. 
Interpreting $P_{gap}(\Delta\eta)$ as a gap probability 
distribution and renormalizing it to unity by dividing it by its integral 
over all phase space~\cite{R,corfu} yields excellent agreement with the 
all data (see Fig.~\ref{fig:softdata}).

\vspace*{-0.5in}
\begin{minipage}[t]{0.5\textwidth}
\phantom{xxx}
\hspace{-0.2in}\includegraphics[width=0.88
\textwidth,height=1.09\textwidth]{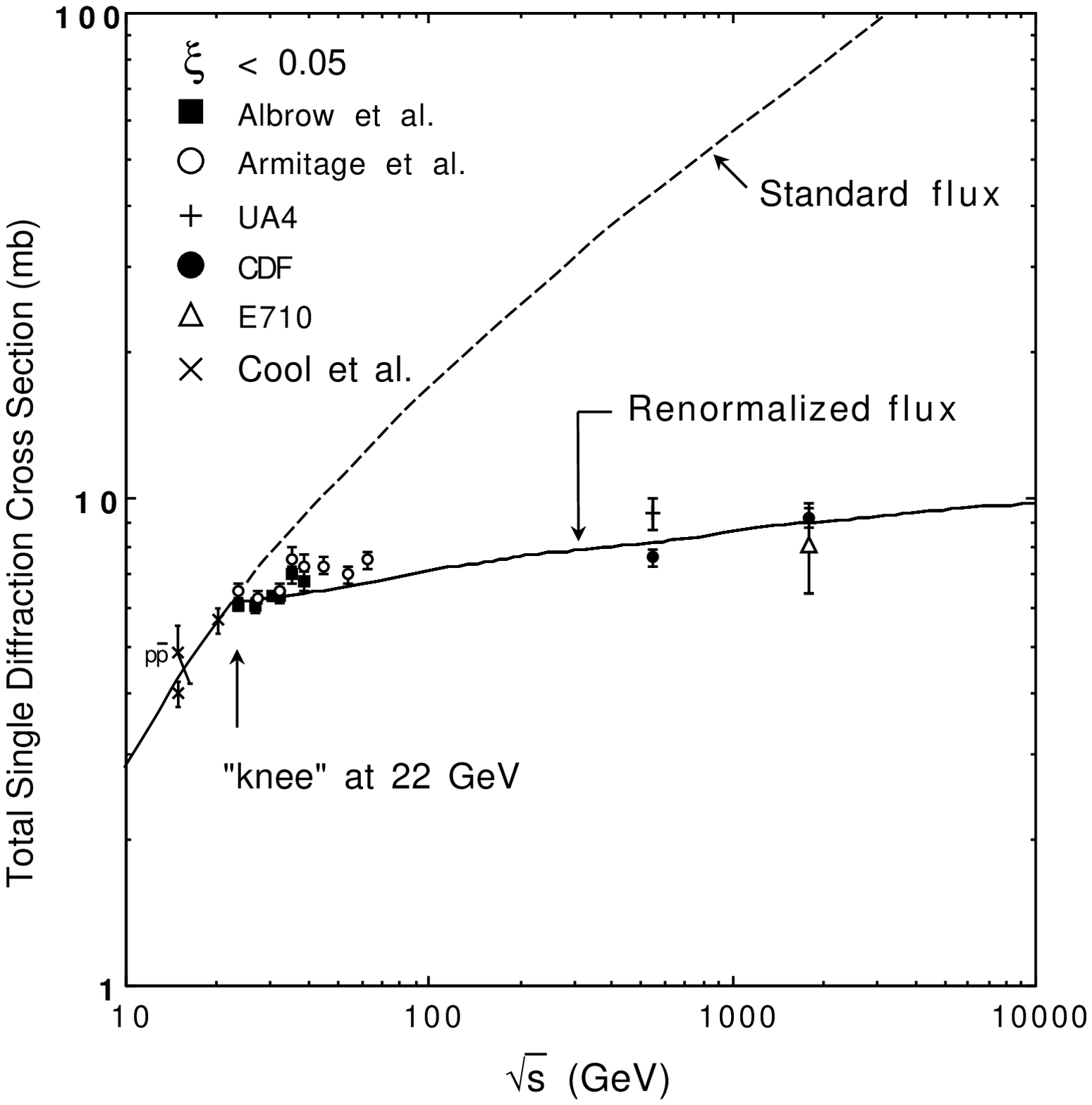}
\end{minipage}
\hspace{0.0in}
\begin{minipage}[t]{0.5\textwidth}
\phantom{xxx}
\vspace*{0.27in}
\hspace*{0.05in}\includegraphics[width=0.75\textwidth]{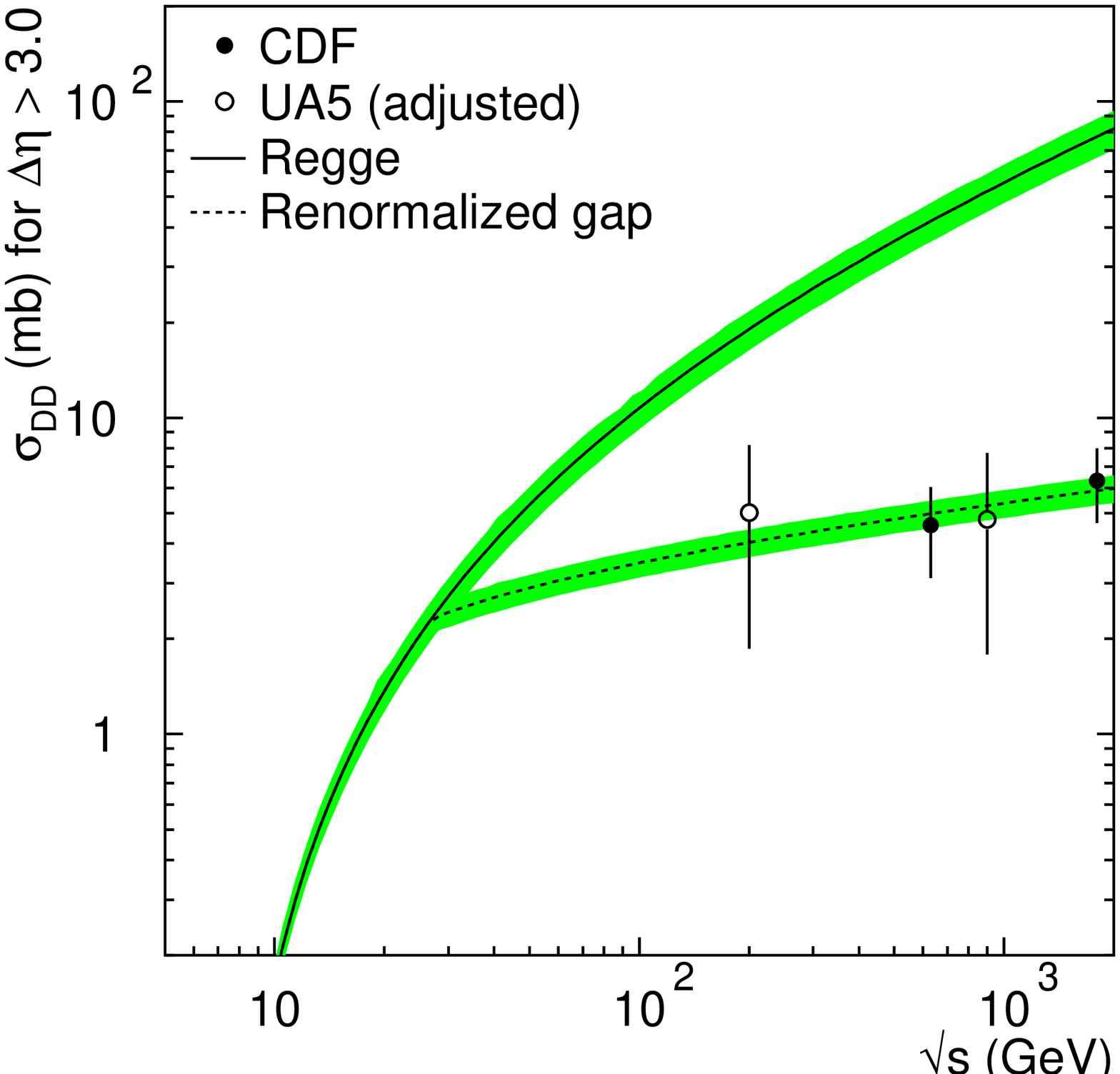}
\end{minipage}

\vglue -0.8in
\begin{minipage}[t]{0.5\textwidth}
\phantom{xxx}
\includegraphics[width=0.75\textwidth]{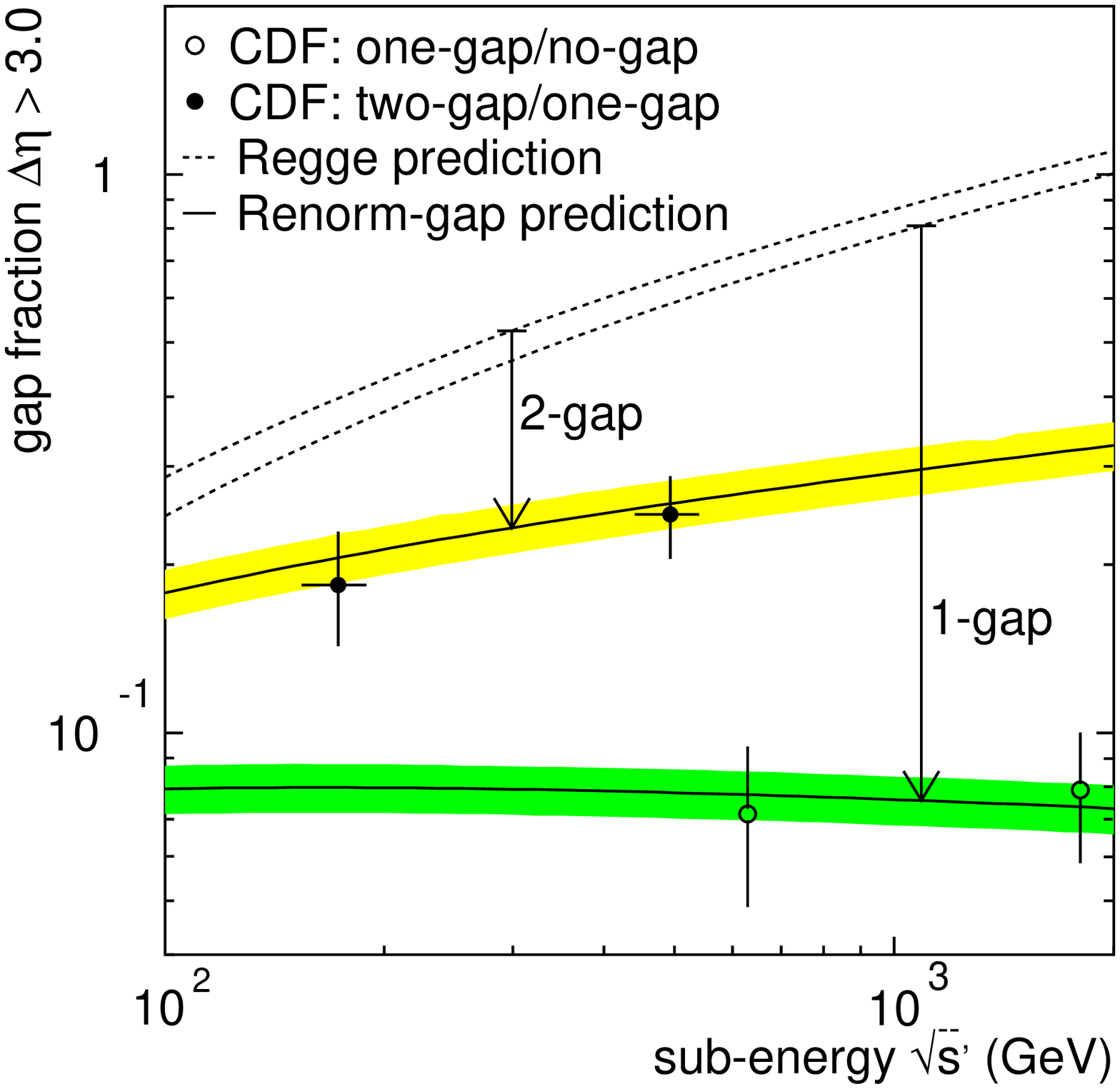}
\end{minipage}
\hspace{0.0in}
\begin{minipage}[t]{0.5\textwidth}
\phantom{xxx}
\vspace*{0.44in}
\includegraphics[width=0.79\textwidth]{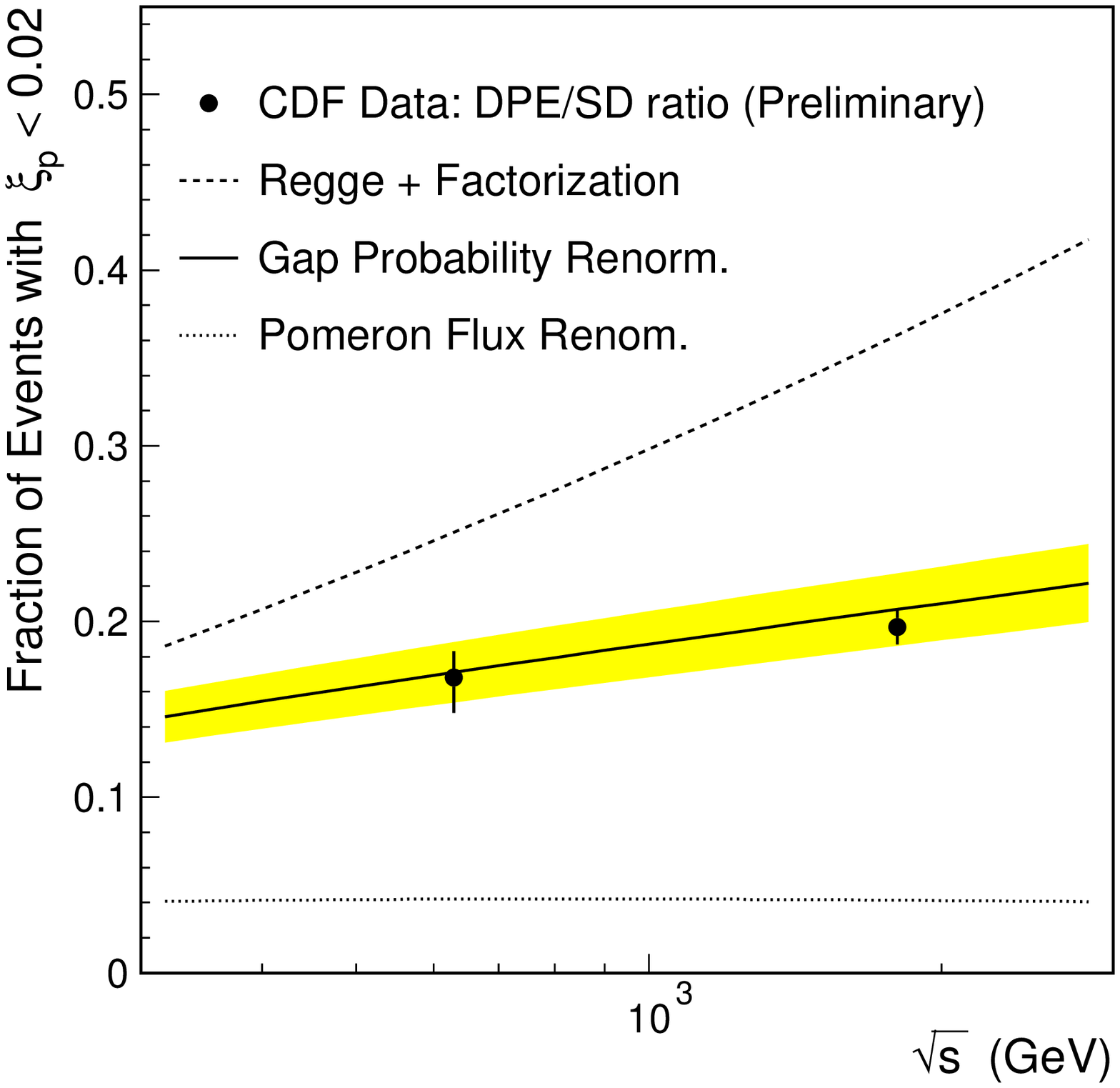}
\end{minipage}

\vglue -0.8in
\begin{figure}[h]
\caption{Soft diffraction cross sections compared with Regge theory 
and RENORM model predictions: (a) SD, (b) DD, (c) ratios of SDD to SD and 
DD to TOTAL, (d) ratio of DPE to SD.}
\label{fig:softdata}
\end{figure}

\vglue -4.5in\hspace{2.2in}{(a)}
\vglue 1.99in\hspace{2.2in}{(c)}
\vglue -2.36in\hspace{5.25in}{(b)}
\vglue 2.00in \hspace{5.25in}{(d)}
\vglue 2.35in

\clearpage
\section{HARD DIFFRACTION}
Hard diffraction processes studied by CDF include SD (dijet, $W$, $b$-quark and
$J/\psi$), DD (dijet) and DPE (dijet) production,
corresponding to the topologies shown in Fig.~\ref{fig:hard}. 
\includegraphics[width=1.0\textwidth]{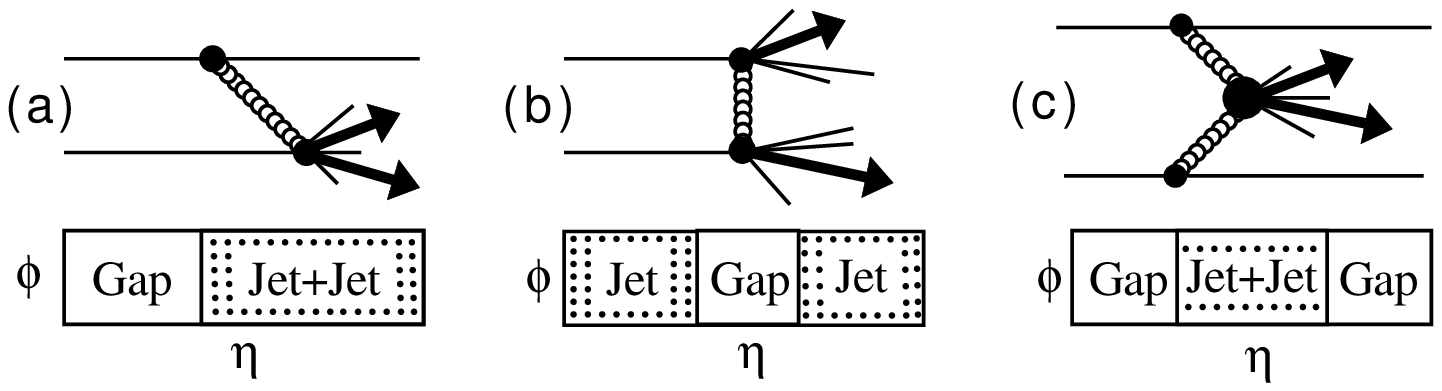}
\vspace*{-6.7in}
\begin{figure}[h]
\caption{Topologies in $\eta$-$\phi$ space for hard (a) SD, (b) DD and 
(d) DPE processes.}
\label{fig:hard}
\end{figure}

Two types of results have been obtained: 
diffractive to non-diffractive cross 
section ratios (using the rapidity gap signature 
to select diffractive events), 
and diffractive to non-diffractive structure function ratios (using a 
Roman Pot Spectrometer to trigger on leading antiprotons).  
For a recent review of CDF Run~I hard diffraction 
result see Ref.~\cite{perspective}. Here
we summarize the aspects of the Run~I results that point to the QCD structure of 
the Pomeron and present new results from Run~II.

\paragraph{Run I rapidity gap results}
(a) At $\sqrt s=$1800 GeV, the SD/ND ratios  
for dijet, $W$, $b$-quark and $J/\psi$ production, as well the ratio of
DD/ND dijet production, are all $\approx 1\%$.
This ``gap fraction'' is suppressed relative to QCD inspired theoretical 
expectations ({\em e.g.} 2-gluon exchange) 
by a factor of $\sim$10, which is comparable to the suppression 
factor observed in soft diffraction relative to Regge theory expectations 
based on factorization.
(b) The gluon fraction of the diffractive exchange was determined 
from dijet, $W$ and $b$-quark production to be $0.54\pm 0.15\%$,
which is similar to the ND fraction.\\
The above results indicate 
that (i) the diffractive structure function is similar to the 
ND one,  apart from an overall suppression in 
normalization, and (ii) at fixed $\bar{p}p$ collision energy QCD 
factorization approximately holds within the diffractive sector.  
\paragraph{Run I Roman Pot Results} 
(a) The diffractive structure function determined from SD dijet production 
is suppressed by a factor of $\sim$10 relative to expectations based 
on extrapolations from parton densities determined from diffractive DIS 
at HERA. This suppression 
is approximately the same as that observed in soft diffraction.
(b) The ratio of SD to ND structure functions behaves 
approximately as $x_{Bj}^{-0.5}$. For a
prediction of such behavior by the RENORM model see Ref.~\cite{newapproach}.
(c) The double-ratio of (DPE/SD)/(SD/ND) structure functions was found to be
$5.3\pm 2.0$, which is equal within errors to the ratio of 
(two-gap/one-gap)/(one-gap/no-gap) in soft diffraction (Fig.~\ref{fig:hard}c).

\paragraph{Conclusions from Run I results}
In both soft and hard diffraction processes cross sections factorize into 
two terms, one containing the cross section at the sub-energy of the 
diffractive cluster and another representing the gap probability distribution,
which must be normalized to unity. A color factor is required for each gap. 
Diffraction appears as the interaction between low-$x$ partons 
subject to color-matching constraints imposed by the rapidity gap 
requirement, as prescribed by the RENORM model.

\paragraph{Run II results}
In Run II, diffractive data have been collected by CDF at $\sqrt s=1.96$ 
TeV and results obtained 
on the $Q^2\equiv (E_T^{jet})^2$ dependence of the diffractive structure 
function and on exclusive dijet production in hard DPE. Results are shown 
in Fig.~\ref{fig:run2}.

\hspace*{-0.2in}\begin{minipage}[t]{0.5\textwidth}
\phantom{xxx}
\includegraphics[width=1.0\textwidth]{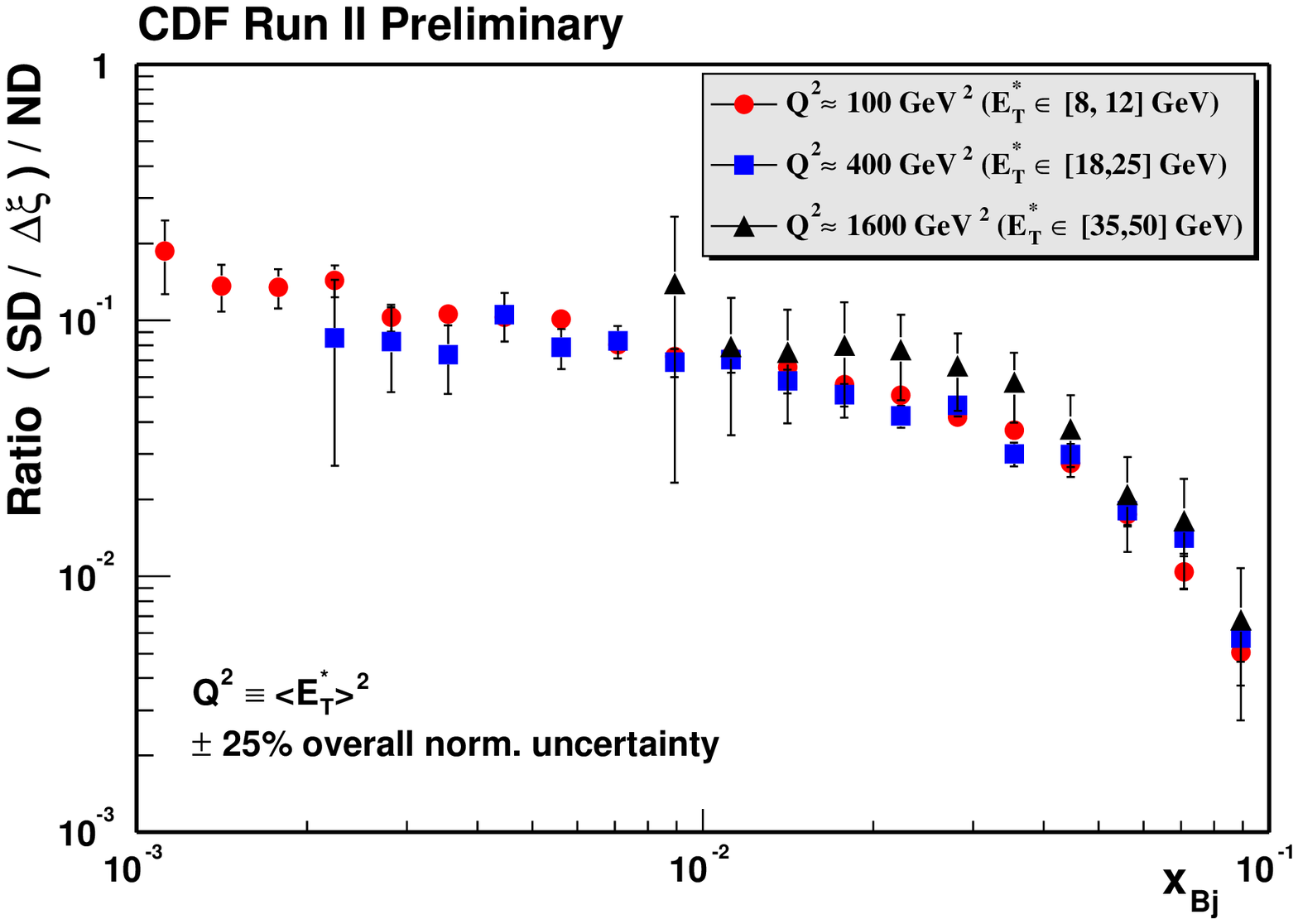}
\end{minipage}
\hspace{0.0in}
\begin{minipage}[t]{0.5\textwidth}
\phantom{xxx}
\includegraphics[width=1.08\textwidth]{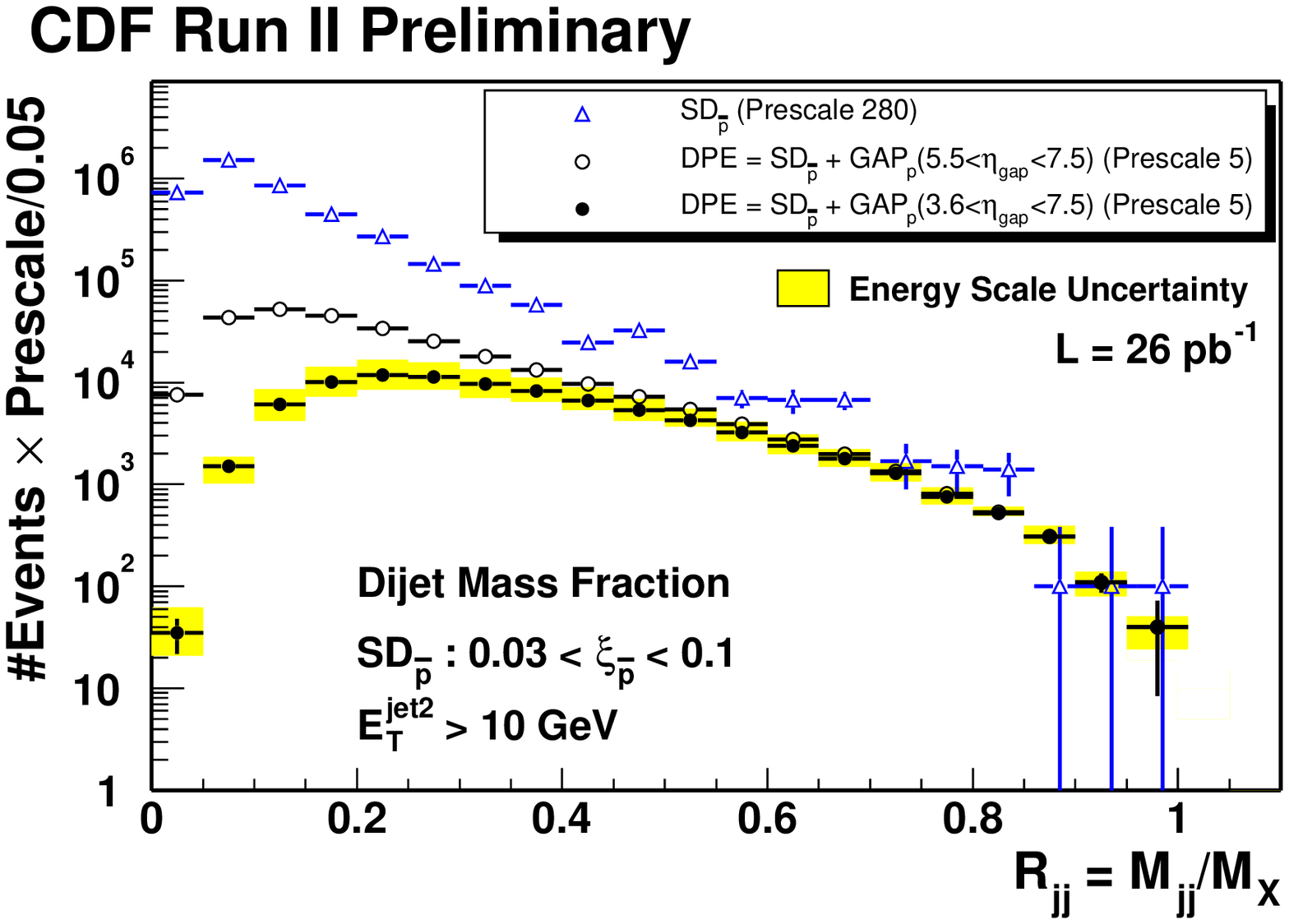}
\end{minipage}

\vglue -0.05in
\begin{figure}[h]
\caption{({\em left}) Ratio of SD/$\Delta\xi_{\bar{p}}$
over ND rates  obtained
from dijet data at various $Q^2$ ranges; {(\em right}) ratio of dijet mass to 
total mass ``visible'' in the calorimeters for dijet production in events 
with a leading antiproton within $0.3<\xi_{\bar{p}}<0.1$ and various 
gap requirements on the proton side: ({\em triangles}) no gap requirement, 
({\em open circles}) gap in 
$5.5<\eta<7.5$, and ({\em filled circles}) gap in $3.5<\eta<7.5$.}
\label{fig:run2}
\end{figure}
 
The ratio of SD/ND rates, which in LO QCD is equal to the 
ratio of the corresponding structure functions at a given $x_{Bj}$, 
shows no appreciable $Q^2$ dependence. 
This result supports the RENORM model, in which the 
diffractive structure function is basically {\em extracted} from the 
non-diffractive one.

Exclusive dijet production in DPE, which has been proposed as a process on 
which to {\em calibrate} models of diffractive Higgs 
production~\cite{KMR}, would appear in 
Fig.~\ref{fig:run2} as a peak in the vicinity of $R_{jj}=1$.
No such peak is observed in the data. For dijets of minimum $E_T^{jet}$
of 10 GeV [25 GeV], the cross section for $R_{jj}>0.8$ is measured 
to be $970\pm65{\rm (stat)}\pm272{\rm (syst)}$ [$34\pm5\pm10$] pb.
Although similar values are obtained in 
Ref.~\cite{KMR} for exclusive dijets, we emphasize that no exclusive 
signal is seen in the data.

\end{document}